\documentclass[10pt,conference,final,finalsubmission,twocolumn]{IEEEtran}

 \usepackage[utf8]{inputenc}
 \usepackage{amsmath}
 \usepackage{amsfonts}
 \usepackage{amssymb}
 \usepackage{graphicx}
 \usepackage{color}
 \usepackage{blindtext}
 \usepackage{scalefnt}
 \usepackage{bbm}

 \usepackage{algorithm}
\usepackage{algorithmic}

 \def\n{N}

\def\hL{\tilde{\Lambda}}

 \def\fraction{\nu}

 \def\rate{{R}}

 \def\rateC{\rate_{J,K,\nu}}

\usepackage{tikz}
\usetikzlibrary{fit,positioning}
\usetikzlibrary{shapes,matrix,decorations.markings,arrows}
\usetikzlibrary{graphs}
\usepackage{pgfplots}
\usepackage{pgfplotstable}
\usepackage{filecontents}
\usetikzlibrary{plotmarks}
\usetikzlibrary{positioning}
\usepgfplotslibrary{fillbetween}

\usetikzlibrary{external} 
\usepackage[left=1.5cm,top=1.5cm,right=1.5cm,bottom=1.5cm]{geometry}
\title{On Generalized LDPC Codes \\for 5G Ultra Reliable Communication\vspace{-2mm}}

\author{
\IEEEauthorblockN{Yanfang Liu$^{*}$, Pablo M. Olmos$^{*}$, and David G. M. Mitchell$^\dag$\\}
\IEEEauthorblockA{
	$^{*}$Universidad Carlos III de Madrid \& Gregorio Mara\~n\'on Health Research Institute, Madrid, Spain\\
	\texttt{\{vivian,olmos\}@tsc.uc3m.es} \\
	$^\dag$Klipsch School of Electrical and Computer Engineering, New Mexico State University, Las Cruces, USA\\
	\texttt{dgmm@nmsu.edu}\vspace{-6mm}
	}
	
	\thanks{
}
}

\usepackage{pdfpages}

\begin{document}

\maketitle

\begin{abstract}
Generalized low-density parity-check (GLDPC) codes, where single parity-check (SPC) constraint nodes are replaced with generalized constraint (GC) nodes, are a promising class of codes for low latency communication. In this paper, a practical construction of quasi-cyclic (QC) GLDPC codes is proposed, where the proportion of generalized constraints is determined by an asymptotic analysis. We analyze the message passing process and complexity of a GLDPC code over the additive white gaussian noise (AWGN) channel and present a constraint-to-variable update rule based on the specific codewords of the component code. The block error rate (BLER) performance of the GLDPC codes, combined with a complementary outer code, is shown to outperform a variety of state-of-the-art code and decoder designs with suitable lengths and rates for the 5G Ultra Reliable Communication (URC) regime over an additive white gaussian noise (AWGN) channel with quadrature PSK (QPSK) modulation. 
\end{abstract}


\section{Introduction}
In order to support an extremely high user density, as well as numerous device-to-device and machine communications, fifth-generation (5G) systems aim to increase the capacity of existing mobile networks by a factor of 1000  \cite{chen2014requirements}. The proposed Ultra Reliable Communication (URC) regime of 5G constitutes a critical component to achieve such goal, as it will enable low-cost and power-efficient anywhere and anytime signalling services \cite{popovski2014ultra}. A number of potential candidate codes for 5G URC have been proposed recently. A representative summary can be found in \cite{sybis2016channel}, where the authors compared turbo, low-density parity-check (LDPC), polar, and convolutional codes, both in terms of performance and computational decoding complexity. To meet the tentative constraints of machine-to-machine (M2M) communications, the authors in \cite{sybis2016channel} consider a low coding rate $R = 1/12$ and short block length (480 bits or 2400 bits). A polar code stands out in the performance comparison, although this solution is limited by the decoding delay imposed by the the sequential nature of successive cancelation (SC) decoding algorithms, which ultimately limits the decoding throughput.

%

Generalized LDPC (GLDPC) block codes were first proposed by Tanner \cite{Tanner81}, and are constructed by replacing some/all of the single parity-check (SPC) constraint nodes by more powerful generalized constraint (GC) nodes, where the GC nodes can be any $(n,k)$ linear code and the $n$ input bits to the GC node are checked correspondingly. The sub-code associated to each GC node is referred to as a \emph{component code}. GLDPC codes have many potential advantages compared to conventional LDPC codes, including fast convergence speed, improved performance in noisy channels, and low error floors \cite{Liva08,Mitchell13GLDPC}. Examples of component codes used in the literature are Hamming codes \cite{Lentmaier99}, Hadamard codes \cite{Yue07}, and expurgated random codes \cite{Liva08,Paolini10}. 

In this paper, we propose a novel GLDPC design methodology that has its roots in a recent contribution by some of the authors in \cite{DBLP:journals/corr/abs-1709-00873}. Here it is shown that, as we vary the proportion of GC nodes in the GLDPC code graph, the tradeoff between rate and iterative decoding threshold presents a unique optimal operational point where the gap to capacity is minimized. Using a GLDPC code operating at exactly this rate (typically larger than the $R=1/12$ target rate), we first optimize a quasi-cyclic (QC) graph lifting to avoid the dominant error events we observe by testing randomly constructed codes. The QC structure also has the benefit of efficient hardware implementation and analysis \cite{li2006efficient, fossorier2004quasicyclic}. We find that the error control performance is somewhat sensitive to the placement of the GC nodes and perform a search for a good distribution of GC nodes, avoiding weak areas. Next, in order to achieve the desired target rate, we propose to combine the optimized GLDPC code with an outer hard-decision decoded, low-complexity code that is designed to match the overall rate to $R=1/12$. In addition to achieving the desired rate, we show that the outer code allows a good performance/complexity trade-off, since the GLDPC code can achieve good enough performance for the outer code to clean up the remaining errors. To the best of our knowledge, the idea of combining a GLDPC code with an outer code as a viable solution for 5G URC is novel in the area. 

In particular, we propose an exemplary design that combines a protograph-based $(2,6)$-regular GLDPC code, in which $75\%$ percent of the nodes correspond to (6,3) shortened Hamming component codes and $25\%$ are SPC nodes, with a rate $R=1/2$ outer code, e.g., a Bose-Chaudhuri-Hocquenghem (BCH) code. Protographs \cite{Thorpe03} impose a structure on the derived code graph, which facilitates the design of fast decoders and efficient encoders, as well as a refined control on the derived graph edge connections \cite{liva2007protograph}. We note also that, unlike a conventional LDPC code, a $(2,6)$-regular GLDPC code has good distance properties and message passing performance \cite{liva2007protograph}. Using such a simple regular graph with a QC lifting allows an efficient implementation of the GLDPC message passing decoder, since variable nodes only have to propagate incoming messages without performing any computation. We provide a summary of the GLDPC decoding complexity, and demonstrate that relatively few iterations of the GLDPC code are required to achieve acceptable error control performance - a critical feature for severely power-constrained devices utilizing 5G URC. Finally, we compare the code performance with the candidates proposed in \cite{sybis2016channel} over an AWGN channel with QPSK modulation showing favorable results.

\section{Background}\label{GLDPC_code}

In this section, we introduce the notation used to define the properties of the GLDPC code ensembles considered in this paper. Throughout, we consider $(J,K)$-regular graphs, where $J$ is the variable node degree and $K$ is the check node degree. Regular graphs are attractive for VLSI decoder implementation and possess robust finite-length scaling behavior \cite{Urbanke09}. 
\begin{figure}[t]
\centering
\includegraphics[scale=1.0]{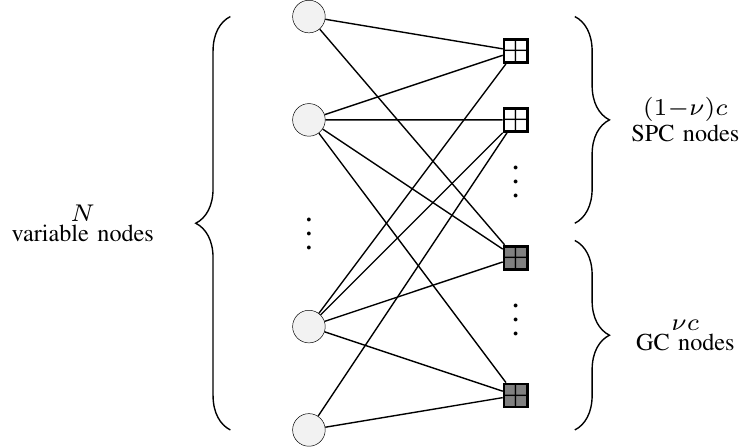}
\caption{Tanner graph of a GLDPC code.}\label{GLDPCgraph}
\end{figure}

Following \cite{DBLP:journals/corr/abs-1709-00873}, we consider a GLDPC code ensemble that is obtained from an LDPC code ensemble (e.g., an LDPC code ensemble defined by a protograph \cite{Thorpe03} or following an degree distribution $(\lambda(x),\rho(x))$) by replacing a fraction $\fraction$  of SPC nodes with identical GC nodes corresponding to an $(n,k)$ component code, while the remaining constraint nodes are SPC. The Tanner graph of a GLDPC code from such an ensemble with block length $\n$ is illustrated in Fig. \ref{GLDPCgraph}. The Tanner graph of any code in this ensemble contains  $\n$ variable nodes, $\fraction\frac{J}{K}\n$ GC nodes, and $(1-\fraction)\frac{J}{K}\n$ SPC nodes. The LDPC ensemble obtained by taking $\fraction=0$ is referred to as the \emph{base LDPC code ensemble} or simple the \emph{base ensemble}. The design rate of the base ensemble $\rate_0$ is given as $\rate_0=1-J/K$ and the design rate $\rateC$ of the GLDPC ensemble is given by $\rateC=\rate_0-\fraction(1-\rate_0)(k-1)$. 
We assume that the incoming edges to every degree $K$ GC node are assigned uniformly at random to each position of the component code. Using the asymptotic analysis proposed in \cite{DBLP:journals/corr/abs-1709-00873} for a binary erasure channel (BEC), we investigate the tradeoff between rate and the threshold as a function of $\fraction$ for a $(2,6)$-regular protograph-based GLDPC code ensemble with a fraction $\fraction$ of rate $R = 1/2$ shortened $(6,3)$ Hamming linear block codes as  GC component codes.\footnote{The protograph of the base ensemble is defined by the all-ones \emph{base matrix} $\mathbf{B}$ of size $2\times 6$, and base ensemble obtained following the usual permutation matrix based protograph construction method, see \cite{Thorpe03}.} Note that this tradeoff is optimized when $\fraction = 0.75$ at the point where the gap to capacity is minimized, resulting in a coding rate of $\rateC = 1/6$. 
Other regular ensembles with higher densities can be explored, but as a representative example in this paper we will restrict our analysis to the $(2,6)$-regular case. This ensemble is of practical interest since it minimizes the GLDPC decoding complexity and simplifies message passing as a result of its low density. 

\begin{figure}
\centering
\includegraphics[scale=1.0]{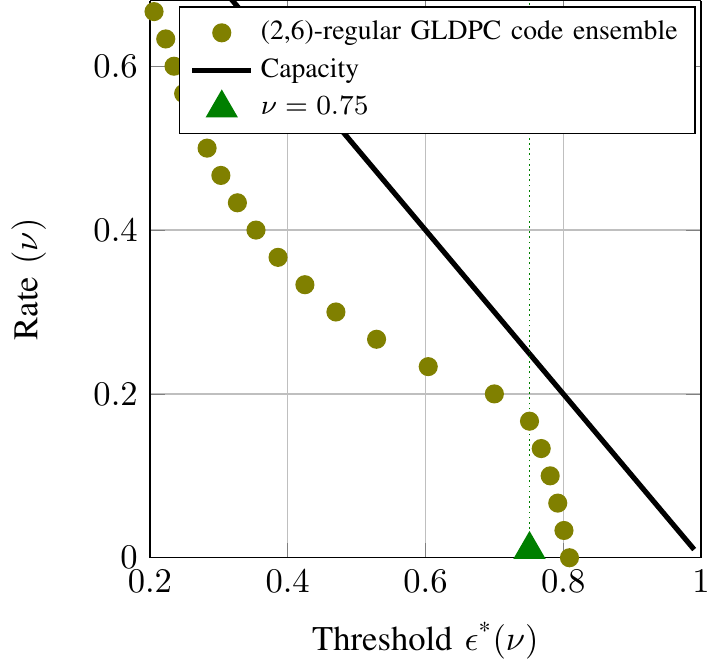}
\caption{Design rate vs. BER threshold of an ensemble of $(2,6)$-regular GLDPC codes as a function of the proportion of GC nodes $\fraction$.}\label{gldpc_26}
\end{figure}

\section{Practical GLDPC Code Design for 5G URC}\label{design_process}

In this section, we investigate several aspects of code design, including QC lifting, placement of GC nodes, and outer code design/code rate matching. 

\subsection{QC Graph Lifting}
The base LDPC code ensemble can be drawn from a random ensemble defined by a degree distribution $(\lambda(x), \rho(x))$, from a semi-structured protograph-based ensemble, or from the structured sub-ensemble of QC codes, where the permutation matrices selected in the protograph-based construction are restricted to be circulant. It is well known  that the algebraic structure of QC codes allows simple encoding using shift registers, with a complexity linear in the block length \cite{costello1983error}. Properly-designed QC graphs have been shown to perform as well as computer-generated random LDPC codes, regular or irregular, in terms of bit-error performance, block-error performance, and error floor for codes with short to moderate block lengths \cite{li2006efficient}. 

Following \cite{fossorier2004quasicyclic}, we first write the parity-check matrix $\bf H$ of a $(J,K)$-regular QC-LDPC code lifted from the all-ones base matrix $\mathbf{B}$ of size $J\times K$ with lifting factor $s$ and code length $ \n = sK$ as follows
\begin{align}\label{H_LDPC}
\mathbf{H} = 
\begin{bmatrix}
I(0)   & I(0)            & \dots & I(0) \\
I(0)   & I(s_{1,1})   & \dots & I(s_{1,K-1})  \\
\vdots & \vdots         &  & \vdots \\
I(0)    & I(s_{J-1,1})  & \dots & I(s_{J-1,K-1})  
\end{bmatrix},
\end{align}

\begin{figure}
\centering
\begin{tabular}{cc}
\hspace{0cm}\includegraphics[scale=1.0]{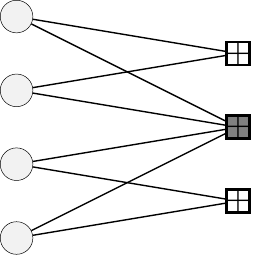} & \hspace{0cm} \includegraphics[scale=1.0]{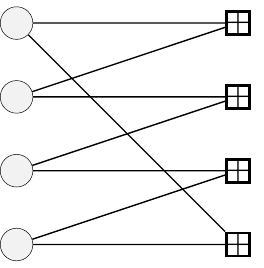}\\
(a) & (b)
\end{tabular}
\caption{Dominant error patterns detected in randomly constructed $(2,6)$-regular QC GLDPC codes.}\label{errorpattern}
\end{figure}

\noindent where $s_{i,j}, 1 \leq i \leq J-1, 1 \leq j \leq K-1$ are the \emph{left shifting parameters}, such that $I(0)$ is the $s\times s$ identity matrix and $I(s_{i,j})$ is the left shifted $s \times s$ identity matrix where each row of $I(0)$ is circularly shifted to the left by $s_{i,j}$ positions. In order to guide our design, we randomly sampled QC codes from the $(2,6)$-regular ensemble and determined empirically the dominant error objects, shown in Fig. \ref{errorpattern}. Structure 1, shown as Fig. \ref{errorpattern}(a), corresponds to two 4-cycles connected by a GC node and Structure 2, shown as Fig. \ref{errorpattern}(b), corresponds to an 8-cycle composed of SPC nodes. Both of these objects, and some other less dominant objects not shown, can be eliminated by increasing the girth $g$ of the base LDPC graph.   

To ensure that the matrix $\bf H$ defined in \eqref{H_LDPC} has a girth of at least $2(i+1)$, a necessary and sufficient condition  \cite{fossorier2004quasicyclic} is $\sum_{t=0} ^{m-1} \Delta_{j_{t}, j_{t+1}} (k_{t}) \neq 0  \bmod  s$,
where $\Delta_{j_{t}, j_{t+1}} (k_{t}) = s_{j_{t},k_{t}} - s_{j_{t+1},k_{t}}$, for all $2 \leq m \leq i$, $0 \leq j_{t}, j_{t+1} \leq J-1$, and $0 \leq k_{t} \leq K-1$, with $j_{0} = j_{m}$, $j_{t} \neq j_{t+1}$, and $k_{t} \neq k_{t+1}$. Given that, in this case study, the target block length is 480 bits, all check nodes have degree $6$ and that $s$ should be chosen to be a prime, we select $s=83$ which gives a slightly larger block length of 490 bits. The resulting $(2,6)$-regular matrix has the form
\begin{align}\label{H_LDPC_26}
\mathbf{H}_{(2,6)} = 
\begin{bmatrix}
I(0)   & I(0)            & I(0)            & I(0)            & I(0)           & I(0) \\
I(0)   & I(s_{1,1})   & I(s_{1,2})   & I(s_{1,3})  & I(s_{1,4})  & I(s_{1,5})  
\end{bmatrix}
\end{align}

There are many possible ways of choosing $s_{1,j}, 1 \leq j \leq 5$. We remark at this point that girth optimization is greatly facilitated by the low-density $(2,6)$-regular structure. Note that $(2,6)$-regular LDPC codes have poor distance properties. In fact, any QC-LDPC code in the form of \eqref{H_LDPC_26} has $d_{min} \leq 6$, independent of $s$ \cite{mackay2001evaluation}. This implies, in turn, that the largest girth we can achieve is $g = 12$ in the base LDPC code, since a cycle of length $2c$ implies the existence of a codeword of weight $c$ in a $(2,K)$-regular code. In order to choose the shift parameters, we make use of Theorem 2.1 in \cite{fossorier2004quasicyclic}, and adopt the so-called $\emph{Power}$ construction to select $s_{1,j}$, $1 \leq j \leq 5$ as $s_{1,j} = a^{1}b^{j} \bmod s$.
For $s = 83$, we can choose $a$ and $b$ as two distinct nonzero elements of $\text{GF}(s)$. When $a =2, b=3$  $[s_{1,1}, s_{1,2},s_{1,3},s_{1,4},s_{1,5} ] = [77, 65, 60, 76, 62]$, resulting in an $\mathbf{H}$ matrix with (optimal) girth $g = 12$. 

\begin{figure}
\centering
\includegraphics[scale=1.0]{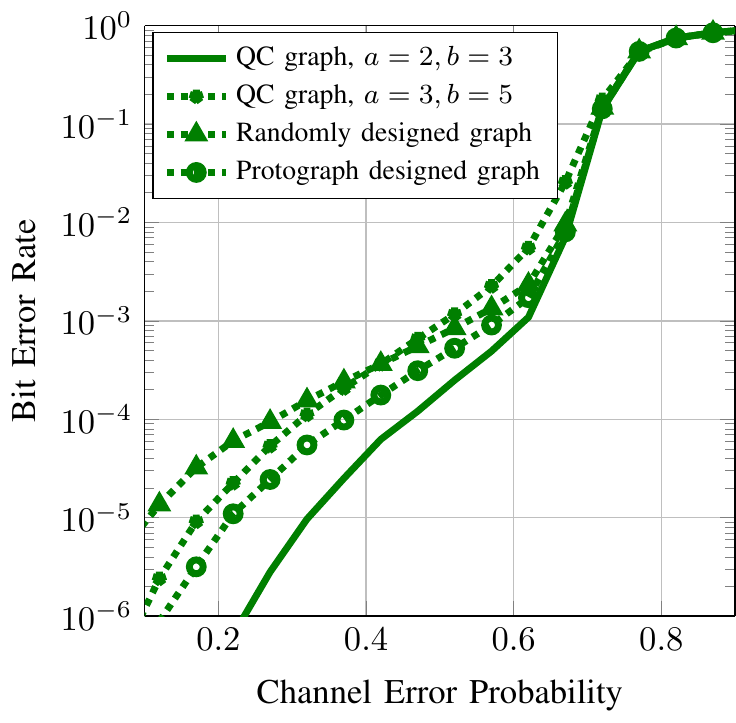}
\caption{Bit error rate for the BEC for a selection of $(2,6)$-regular LDPC codes with different Tanner graph constructions.}\label{gldpc_bec}
\end{figure}

In Fig. \ref{gldpc_bec}, we plot the bit error rate performance on the BEC of several base LDPC codes, including QC-LDPC codes constructed following the power method with $[a,b] = [2,3]$ and $[3,5]$ (with girth $g=6$), a randomly constructed $(2,6)$-regular code, and a randomly constructed semi-structured $(2,6)$-regular protograph-based code. All simulations were allowed $I_{\text{max}} = 100$ iterations. We remark that the waterfall performance of all codes are similar, but the error floor of the optimized QC-LDPC base code is reduced significantly.\footnote{Note that representative curves are plotted here from a selection of simulation results.} We will see later that this property will be inherited by the code after a certain fraction of SPC nodes are replaced by GC nodes.

\subsection{Location of the GC Nodes}
Given the matrix of the base LDPC code, we now turn our attention to the locations of the GC nodes. As discussed in Section~\ref{GLDPC_code}, the optimal proportion, from a threshold perspective, is that $75\%$ percent of the check nodes should be replaced by GC nodes. The performance of the GLDPC codes with $75\%$ shortened $(6,3)$ Hamming codes is somewhat sensitive to the locations of GC nodes. In particular, regions of the graph with multiple local SPC nodes should be avoided, since those constraints have less local error correcting capability. We sampled 300 randomly chosen GLDPC codes and selected the best performing code. 

\subsection{Target Coding Rates}
\begin{figure}
\centering
\includegraphics[scale=1.0]{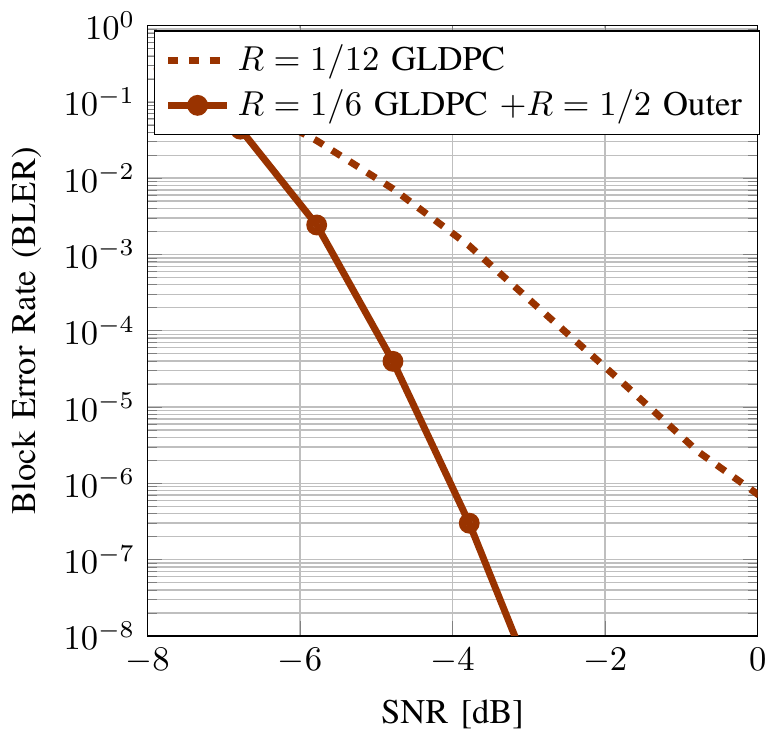}
\caption{BLER on an AWGN channel with QPSK modulation as a function of the SNR for a rate $R=1/12$ GLDPC code and a rate $R=1/6$ GLDPC code with $R=1/2$ 
outer code that can correct 20 errors.}\label{gldpc_16vs112}
\end{figure}
To adapt the designed GLDPC code to the 5G URC regime, we consider techniques to lower the coding rate from $R=1/6$ to $R=1/12$. Among others, this could be achieved by adding more GC nodes and/or utilizing an outer code. Both approaches have advantages and disadvantages. We expect that adding more GC nodes should improve the error correcting performance by improving the local decisions at constraint nodes; however, we know that the threshold of the ensemble will become (relatively) further from the capacity. Alternatively, the outer code can be selected to complement the inner code (error locations/size/type); however, this will require additional circuitry and implementation cost. In Fig. \ref{gldpc_16vs112} we compare the BLER performance of a rate $R = 1/12$ GLDPC code, obtained by selecting $\fraction = 0.875$, versus that of the rate $R = 1/6$ GLDPC code with a rate $R =1/2$ outer code that can correct 20 errors. For this comparison, both $(2,6)$-regular GLDPC codes were randomly constructed following the protograph method with randomly placed GC nodes (similar results are obtained for different random draws of the matrices). The results show that the higher rate GLDPC code, optimized for threshold, with an outer code has significantly better performance than adding more GC nodes. Note that, in addition to good waterfall performance, we do not observe an error floor down to a BLER of $10^{-8}$ with the outer code version. The outer code can be chosen to be any $(n,k)$ linear block code of appropriate length and rate to meet the target, such as a $(83, 40)$ shortened BCH code. We would expect to use a low-cost, high speed, hard-decision decodable code for the outer code. With a block length of 83 bits and a rate $R = 1/2$, we can conservatively assume that the decoder can correct up to 20 errors \cite{costello1983error}.

\section{GLDPC Message Passing}\label{Decoder_awgn}
In this section we discuss the  message passing update rules for the for GLDPC code. Compared to the conventional belief propagation update rules for LDPC decoders, the only difference here is how to process probabilistic messages at the GC nodes. In this regard, this processing completely depends on the chosen component code.
In our example, the generator matrix of the rate $R=1/2$ shortened $(6,3)$ Hamming code is
\begin{align}
\text{G} = 
\begin{bmatrix}
1 &0 &0 &1 &1 &0 \\
0 &1 &0 &1 &0 &1 \\
0 &0 &1 &0 &1 &1
\end{bmatrix}, \label{generator_mat}
\end{align}
and the associated codebook is 
\begin{align}\label{codewords_36hamming}
\text{C} = 
\begin{bmatrix}
0 &0 &0 &0 &0 &0 \\ 
0 &0 &1 &0 &1 &1 \\ 
0 &1 &0 &1 &0 &1 \\ 
0 &1 &1 &1 &1 &0 \\ 
1 &0 &0 &1 &1 &0 \\ 
1 &0 &1 &1 &0 &1 \\
1 &1 &0 &0 &1 &1 \\
1 &1 &1 &0 &0 &0
\end{bmatrix}.
\end{align}

The update rule for a $(6,3)$ Hamming GC nodes is therefore determined by \text{C}. Let $\Lambda_{j}$ denote the input LLR message coming from the $j$-th variable node connected to the GC node, where index $j$, $j=1,\ldots,6$, corresponds to the $j$th input to the component code, and let $\hL_{j}$ denote the output LLR message to be sent to the $j$-th variable node, then it follows that
\begin{align} \nonumber
\hL_{j} &= \text{log}\left[\sum_{\substack{i\in\{1,8\}\\C_{i,j} =0}}\exp\left(\sum_{\substack{m\in\{1,6\}\\m \neq j}} \mathbbm{I}[(C_{i,m} = 0)] (\Lambda_{pj}-\Lambda^*)\right)\right]
\\\label{check_lambda}
&- \text{log}\left[\sum_{\substack{i\in\{1,8\}\\C_{i,j} =1}}\exp\left(\sum_{\substack{m\in\{1,6\}\\m \neq j}} \mathbbm{I}[(C_{i,m} = 1)] (\Lambda_{pj}-\Lambda^*)\right)\right],
\end{align}
where $C_{i,m}$ denotes the $m$-th bit of the $i$-th codeword, $i=1,\ldots,8$,  $\Lambda^*=\max_j \Lambda_{j}$,  and we  use the  \emph{log-sum-exp trick} to avoid numerical issues in the evaluation of the exponential terms. Note that at variable nodes and SPC nodes the message passing update rules those for standard LDPC decoding \cite{kschischang2001factor}.

\section{Decoding Complexity}\label{Decoding_complexity}

We measure the computational complexity  of the decoder by enumerating the number of additions, subtractions, multiplications, divisions, comparisons, max (min) operations, and table look-up operations.\footnote{Although these operations are not equivalent from a hardware implementation point-of-view, we adopt this metric to be consistent with the complexity analysis of coding schemes in \cite{sybis2016channel}.} Most of these operations correspond to one equivalent addition, whereas the comparison operation, in most cases, corresponds to two equivalent additions \cite{sybis2016channel}. In the following, we ignore the hard-decision decoding complexity of the outer code, as the additional complexity is negligible compared to the GLPDC message passing complexity.

According to \eqref{check_lambda}, there are $27\times K$ additions/subtractions and $10\times K$ multiplications/divisions to update every GC node and every SPC node respectively. Furthermore, note that the variable node degree is $J=2$, hence there is only one addition to perform when updating the variable nodes and thus the decoding complexity per iteration for variable node is $J\times N =  996$. Altogether, the decoding complexity per iteration is $J \frac{\n}{K} \fraction 27 K + J \frac{\n}{K} (1-\fraction) 10 K + JN= J \n ( 11 + 17\fraction) = 23655$, given $J =2,  \n = 498$, and $\fraction = 0.75$. If $I_{\text{max}}$ denotes the maximum iteration number, the decoding complexity (in the worst case) is $23655\times I_{\text{max}}$. To minimize $I_{\text{max}}$, we compare the decoding performance for different values as shown in Fig. \ref{compare}. Even with only 5 decoding iterations, the performance degradation compared to the case $I_{\text{max}}=50$ is approximately 0.25 dB at a BLER of $10^{-5}$, and decreases further to approximately $0.05$ dB if we allow 10 decoding iterations. Note that the number of iterations is also related to the decoding throughput.

\begin{figure}
\centering
\includegraphics[scale=1.0]{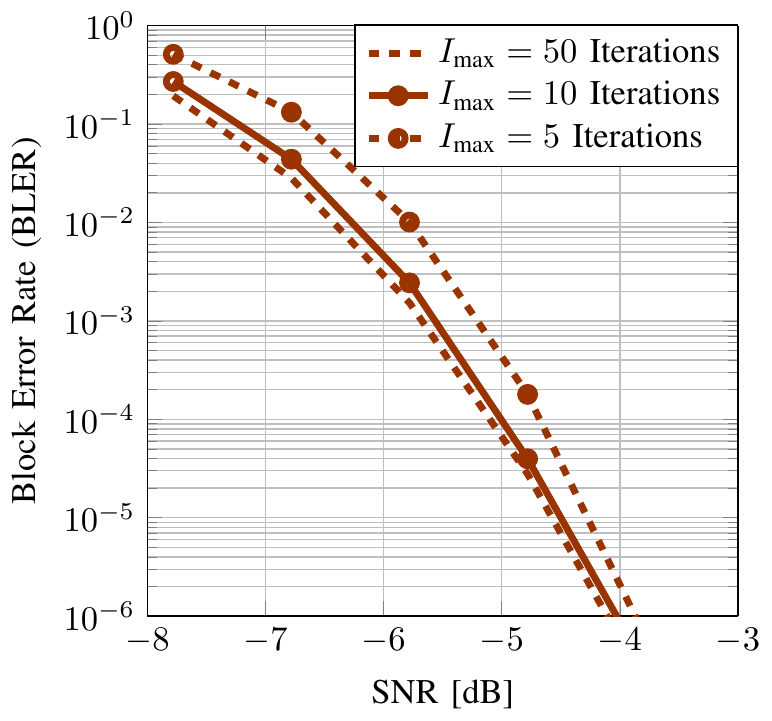}
\caption{BLER of a $(2,6)$-regular GLDPC code with different maximum allowable number of iterations.}\label{compare}
\end{figure}

\section{Experimental Results}\label{results}

In \cite{sybis2016channel}, the authors investigated turbo, LDPC, polar, and convolutional codes with a variety of decoders as candidates for 5G URC. 
To validate the proposed approach, we compare the simulated decoding performance of the $(2,6)$-regular QC GLDPC code with an $R=1/2$ outer code against the aforementioned codes from \cite{sybis2016channel} in terms of BLER performance over the AWGN channel. All coding schemes have 40 information bits and rate $R=1/12$. Fig. \ref{compare} displays the results where the GLDPC decoder was allowed $I_\text{max} = 10$ iterations. We remark again that we use a concatenated scheme (GLDPC and outer code) to achieve improved performance for the targeted coding rate. Two example outer codes are shown: the first only decodes up to 7 errors, but we observe that the performance is comparable with the polar code scheme; the second corrects up to 20 errors and we observe significantly improved performance, close to a 2 dB gain at a BLER equal to $10^{-5}$ over the polar code scheme. Further comparisons with other potential coding schemes, including a variety of polar codes, are the subject of ongoing work, however the results in Fig. \ref{compare} demonstrate the potential of the proposed design methodology: an inner GLDPC code optimized asymptotically for threshold and proportion of GC nodes with finite length QC design based on eliminating problematic objects along with a relatively simple, off-the-shelf, hard-decision decoded outer code that cleans up the remaining errors. With such a performance gain, we believe our proposed design approach is a strong candidate for future URC standards. 

\begin{figure}
\centering
\includegraphics[scale=1.0]{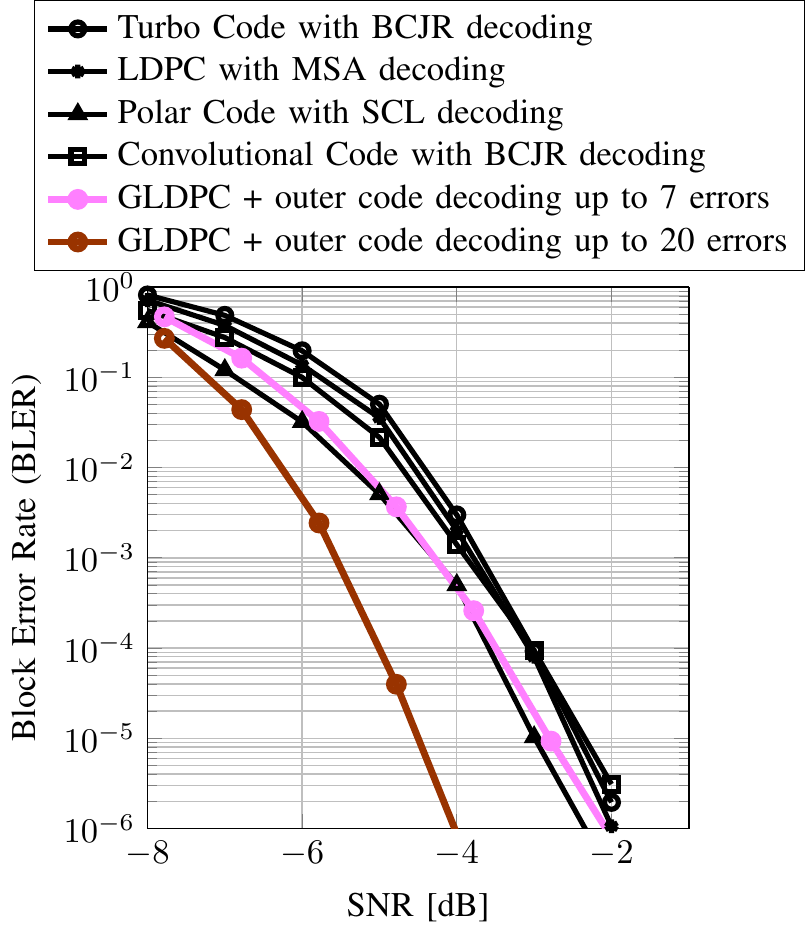}
\caption{BLER over an AWGN channel with QPSK modulation for the proposed coding scheme (GLDPC with outer code). The following coding schemes are also included (details can be found in \cite{sybis2016channel}): turbo code, LDPC with min-sum approximation (MSA) decoder, polar code with successive cancelation list (SCL) decoding, and a convolutional code with Bahl-Cocke-Jelinek-Raviv (BCJR) decoding. All coding schemes have 40 information bits and rate $R=1/12$.}\label{compare}
\end{figure}

\section{Conclusion and Future Work}\label{conclusion}
In this paper, we presented a novel coding scheme for 5G URC based on combining an inner GLDPC code with a simple outer hard-decision decoded outer (e.g., BCH) code. Asymptotically, the proportion of the GC nodes is optimized for threshold while, for good finite-length performance, the GLDPC code is constructed with a simple regular quasi-cyclic graph which is attractive for analysis and VLSI implementation. 
Our results demonstrate that we can achieve remarkable gains compared to existing schemes in the literature and the approach  could be considered for future URC communication standards.

Throughout the paper, a $(2,6)$-regular GLDPC code was used as an example, since it has a number of desirable features in practice, but future work will include a study of ensembles of higher densities, investigation of performance with higher order modulation, and a performance analysis when the numerical precision of the proposed message passing scheme is limited. Further, doubly-generalized LDPC codes will be also explored, since they can provide a large flexibility of code design. 
\section{Acknowledgements}
This work has been funded in part by the European Research Council (ERC) through the European Unions Horizon 2020 research and innovation program under Grant 714161, by the Spanish Ministerio de Econom\'ia y Competitividad and the Agencia Espa\~nola de Investigaci\'on under Grant TEC2016-78434-C3-3-R (AEI/FEDER, EU) and by the Comunidad de Madrid in Spain under Grant S2103/ICE-2845. This work was funded in part by the National Science Foundation under Grant No. ECCS-1710920. 

 \bibliographystyle{IEEEtran}

\end{document}